\documentstyle[multicol,aps,prl,epsf,rotate]{revtex}

\date{\today}
\draft
\tighten
\begin{document}
\def\sqr#1#2{{\vcenter{\hrule height.3pt
      \hbox{\vrule width.3pt height#2pt  \kern#1pt
         \vrule width.3pt}  \hrule height.3pt}}}
\def\square{\mathchoice{\sqr67\,}{\sqr67\,}\sqr{3}{3.5}\sqr{3}{3.5}}
\def\today{\ifcase\month\or
  January\or February\or March\or April\or May\or June\or July\or
  August\or September\or October\or November\or December\fi
  \space\number\day, \number\year}

\def\Bbb{\bf}


\newcommand{\ww}{\mbox{\tiny $\wedge$}}
\newcommand{\pp}{\partial}

\title{Twistors and Holography\thanks{email: chamblin@prancer.physics.louisville.edu}}
\author{Andrew Chamblin }
\address{Theory Division, T-8, Los Alamos National Laboratory,
Los Alamos, NM 87545, USA \\
\& \\
Department of Physics, University of Louisville, Louisville, KY 49292  \\}
\maketitle

\begin{abstract}

We extend Bousso's notion of a lightsheet - a surface where entropy can be defined in 
a way so that the entropy bound is satisfied - to more general surfaces.  Intuitively,
these surfaces may be regarded as deformations of the Bousso choice; in general, these
deformations will be timelike and so we refer to them as `timesheets'.  We show that
a timesheet corresponds to a section of a certain twistor bundle over a given
spacelike two-surface $B$.  We further argue that increasing the entropy flux through
a given region of spacetime corresponds to increasing the volume of certain regions in twistor
space.  Put another way, it would seem that entropy in spacetime corresponds to volume
in twistor space.  We argue that this formulation may point a way towards a version of the
covariant entropy bound which allows for quantum fluctuations of the lightsheet.  
We also point out that in twistor space, it might be possible to give a purely
topological characterization of a lightsheet, at least for suitably simple spacetimes.\\
\\
\end{abstract}

\begin{multicols} {2}

\section{Introduction}
Throughout history \cite{jammer}, one of the central problems faced by philosophers and
scientists has been the simple query: How does one define the concepts of space and
time? Are they merely abstractions which we have introduced in order to facilitate a
description of the inter-relationship between things which actually `exist' (e.g.,
such as material bodies)? Or do space and time `exist' in and of themselves, without
any reference to observable consequences? Although such Machian musings may seem
esoteric, they actually take on a new and exciting life when viewed in the light of
modern ideas coming from theoretical physics, especially quantum gravity. Indeed, most
theoretical physicists today would probably agree that `space' and `time' will be
effective concepts which only emerge at low energies. At scales past the Planck
energy, `space' and `time' will simply cease to have any operational meaning, and some
more fundamental ideas (comprising quantum gravity) will have to take over. 

Of course,
at first it seems nonsensical to assert that the `ultimate theory' should be
constructed without invoking the use of the words `space' or `time'. After all, from
our earliest days of undergraduate physics, we were all weaned on physical theories
which simply would not make any sense without reference to these concepts. More
precisely, theories such as classical or quantum mechanics are useful precisely
because they are bodies of knowledge which allow us to {\it predict}, with at least
some probability, the nature of {\it future} events given some knowledge about {\it
present} or {\it past} events. One of the conceptual obstructions to constructing a
quantum theory of gravity is that it is unclear what the theory will have to do with
prediction; although certain quantum gravity models {\it do} yield predictions (such
as the No Boundary Proposal), much more work on connecting quantum gravity with the
low energy world around us is needed.

On the aesthetic level, however, the construction of a quantum theory of gravity may
be the most beautiful way of solving the puzzle of space and time by showing that the
concepts can be removed altogether from scientific discourse (since they are low
energy manifestations of some more fundamental concepts); surely, Ernst Mach would be
proud of this approach.  On the other hand, even if we do discover such a
`pre-geometric' form of quantum gravity, we will still need to understand how
general relativity can emerge at low energies.

Motivated by this problem, various authors have put forward the idea
that the `Holographic Principle' should somehow be incorporated into any
attempt to construct a quantum theory of gravity.  This principle, which was
first developed in papers by 't Hooft \cite{thooft} and Susskind \cite{lenny},
is on the surface a {\it radical} statement about 
how many degrees of freedom there are in
Nature.  In essence, the principle asserts that a physical system can be
completely described by information which is stored at the boundary of the
system, without exceeding one bit of information per unit Planck area.

For some time, there was no precise covariant statement of the Holographic Principle;
however, this situation was rectified in a series of elegant papers by
Fischler and Susskind \cite{fs} and
Bousso (\cite{raph}, \cite{raph1}, \cite{raph2}).  In particular, by
carefully choosing the lightlike surfaces (called `light-sheets') 
where the entropy of a given
system can reside, Bousso was able to develop a 
mathematically precise covariant entropy conjecture.

Soon after the work of Fischler, Susskind and Bousso (FSB), a proof
of various classical versions of Bousso's bound was provided by
Flanagan, Marolf and Wald (FMW) \cite{fmw}.  In order to make a mathematically
precise statement, which could consequently be proven, they had to take the
crucial step of introducing the notion of an {\it entropy flux vector},
usually denoted $s^a$.  The total entropy through a given light-sheet is then
defined to be the integral of $s^a$ over the surface of the light-sheet.
They showed that FSB-type bounds could be {\it proven},
provided the entropy flux vector satisfied the following two criteria:
\vspace*{0.2cm}

{\noindent A: $(s_a k^a)^2 \le T_{ab} k^a k^b / (16 \pi) + {\sigma_{ab}}{\sigma^{ab}} / 128 {\pi}^{2} $}
\vspace*{0.2cm}

{\noindent B: $|k^a k^b \nabla_a s_b| \le \pi T_{ab} k^a k^b /4 + {\sigma_{ab}}{\sigma^{ab}} / 32 $}
\vspace*{0.2cm}

\noindent where $k^a$ denotes the tangent vector to a given null geodesic $\lambda$
generating the light sheet in question, $T_{ab}$ denotes the stress-energy
tensor, and $\sigma_{ab}$ denotes the shear tensor of the null congruence \cite{he}.  
In essence, one may regard conditions (1) and (2) as the `definition'
of an {\it acceptable} entropy flux vector \footnote{Recently, other simple sufficient 
conditions for the entropy bound have been derived \cite{bfm}.  For the purposes of this
note, and of these conditions will suffice.}.  Crudely, the stress-energy part of 
the entropy flux is generated by `matter' degrees of freedom, whereas the shear part
corresponds to purely gravitational degrees of freedom \cite{josh}.  Finally, we 
point out that there have been some criticisms of the covariant entropy bound and of
the holographic principle more generally, see
e.g., \cite{dave}, \cite{reza}, \cite{lee} and \cite{nem} as examples.

In this paper, we adopt the philosophy that ultimately a quantum theory of gravity
will be some structure which relates {\it information} - in some primal form - with
the geometry of space and time.  A natural question is therefore:  `Where' is information
stored, and how much information can we store there?  The covariant entropy bound gives us a
proposal: Information is stored on those surfaces where the entropy bound holds.  It therefore
behooves us to classify the most general set of such surfaces.

\section{Twistors, timesheets and entropy}

According to one definition, a `null' twistor is a null geodesic in Minkowski space
\cite{ronnie}, \cite{pr} \footnote{For the sake of simplicity, we will use the term 
`twistor' and `null twistor' interchangeably here.}.  
Consider a point $x$ in spacetime.  Then there is a full
`lightcone' worth of null rays through that point.  Put another way, a point in spacetime
corresponds to an $S^2$ in twistor space.  Similarly, consider some spacelike 2-surface
$B$.  Then the space of twistors over B is an $S^2$ bundle over B.  We denote this bundle of
null rays over $B$ $T_B$:
\[
T_{B} = S^{2} ~{\longrightarrow}~ B
\]
Given the bundle of null rays over $B$, we can consider smooth sections of this bundle.
Given a section $s$ of this bundle, we will define certain surfaces, called `timesheets', \footnote{I thank
R. Bousso for suggesting the name `timesheet'.}  
associated with the section $s$ as follows:  Given a point $x$ in $B$, $s$ corresponds to a null
ray or vector.  Since we are assuming the spacetime is time orientable, this can be decomposed
into a future directed geodesic (from $x$), and similarly a past directed geodesic from $x$.
The future timesheet associated with $s$, denoted $T^{+}_{s}(B)$ is obtained by terminating
each future directed null generator of s at any caustic.  There is the obvious corresponding
definition for the past timesheet, $T^{-}_{s}(B)$.  Terminating these surfaces at caustics
is of course the prescription Bousso gives
for terminating lightsheets, the only difference here is that these sections are in general
{\it timelike}.  In order to understand this, it is useful to recall the Raychaudhuri 
equation \cite{he}, \cite{wald}, \cite{roger}:
\begin{equation}
\frac{d{\theta}}{d{\lambda}} = -\frac{\theta^2}{2} - {\sigma}_{ab}{\sigma}^{ab} +
{\omega}_{ab}{\omega}^{ab} - 8{\pi}T_{ab}k^{a}k^{b}
\end{equation}
Bousso assumes that the null generators of lightsheets are everywhere orthogonal to $B$,
which means that the rotation or twist parameter vanishes:
\[
\omega_{ab} = 0
\]
In general one could imagine that the congruence is not orthogonal, and that the twist
is non-vanishing.  Assuming causality, it follows that the surface will in general be
timelike.  Intuitively, the reader should think of d-dimensional Minkowski space as a 
hypersurface in (d+1)-dimensional Minkowski.  It is certainly the case that d-dimensional
Minkowski can be `ruled' by null curves (you just introduce advanced and retarded
null coordinates on the hypersurface), but it is a timelike hypersurface relative to 
the ambient (d+1)-dimensional coordinates.

However, there is one key point here:  In order for the entropy bound to hold,
the expansion parameter $\theta$ must satisfy the crucial inequality
\begin{equation}
\frac{d\theta}{d\lambda} ~{\leq}~ - \frac{\theta^2}{2}
\end{equation}
In particular, (2) must be satisfied everywhere on the timesheet.  Given (1), this means that
the following inequality must hold:
\begin{equation}
8{\pi}T_{ab}k^{a}k^{b} + {\sigma}_{ab}{\sigma}^{ab} ~{\geq}~ {\omega}_{ab}{\omega}^{ab}
\end{equation}
If the inequality (3) is satisfied everywhere along a given timesheet, then we will say
that the timesheet is {\it holographic}, for the simple reason that the entropy bound will
hold on the timesheet.  But there is yet another subtlety: It could be the case
that the twist tensor {\it diverges} too quickly, so that the inequality (3) is violated.
To see that this may not happen, recall the equation for the twist tensor \cite{wald}:
\begin{equation}
k^{c}{\nabla}_{c}{\omega}_{ab} = \frac{d}{d \lambda}({\omega}_{ab}) =  -{\theta} {\omega}_{ab}
\end{equation}
Similarly, the shear tensor satisfies the equation
\begin{equation}
\frac{d}{d \lambda}({\sigma}_{ab}) = -{\theta}{\sigma}_{ab} + \hat{C_{abcd}}k^{c}k^{d}
\end{equation}
where the `hat' ($\hat{C}$) operation over the Weyl tensor $C$ is explained in \cite{wald}.
Now, suppose that the inequality (3) is initially satisfied - then $\theta$ is initially
decreasing towards minus infinity.  If we assume that $\theta$ reaches minus infinity
in a finite amount of affine parameter time, we can estimate how ${\omega}_{ab}$ 
and ${\sigma}_{ab}$ must vary in order for the conjugate point to be reached.  Following along the geodesic,
it follows that as ${\theta}$ runs to minus infinity, ${\omega}_{ab}$ and ${\sigma}_{ab}$ must
`scale' at the same rate, so that their effects are precisely cancelled.  While this does not
forbid violation of (3) at some points where the original covariant entropy bound holds, it does
suggest that as long as there is enough shear it should be possible to allow for fluctuations
away from orthogonality.

So when can holographic timesheets exist?  Well, it is clear that they only exist
when there is a `sufficient' amount of entropy floating around in the spacetime.
If both the shear tensor and the stress-energy tensor vanish, then the rotation
will have to vanish and one must return to using the Bousso prescription.

Turning this around, suppose we try to `force' a given amount of entropy flux through a given
region.  Then the criteria A and B for the covariant entropy bound imply that we must
correspondingly see an increase in either the shear tensor or the stress-energy tensor
through that region.  But this in turn implies that we will have more freedom in our choice
of holographic timesheets.  Put another way, we could think of measuring the `size' of the
space of holographic timesheets.  The space of holographic timesheets on $B$, which we 
denote $H(B)$, is a compact subset of the compact twistor bundle over B (here we are assuming that
$B$ is some compact spacelike 2-surface), so $H(B)$ has some finite volume, vol$H(B)$.
So it would seem that there is a direct relationship between entropy flux in spacetime, and
the volume of the space of holographic timesheet sections in twistor space.
We emphasize that this proposal is a definition of (and indeed a conjecture about)
how entropy in spacetime might be realized in twistor space.

\section{Conclusion: Lightsheets via twistor topology?}

An old dream of the twistor programme was that physics should be formulated in twistor space -
that in some sense twistors are more `fundamental' than spacetime points \cite{pr}, \cite{ronnie}.  Here, we have
explored this possibility, at least for the case of `real' twistors, which may be thought of
as the bundle of null rays over a given spacetime.  Since the modern, fully covariant realizations
of the holographic entropy bound suggest that we should think of information as being defined
on null rays, this is suggestive that twistor theory may indeed play a role in the construction
of a quantum theory of gravity.  At the very least, holography is a little more `obvious' in 
twistor space: Given a holographic timesheet, the entropy integral over the timesheet is manifestly
two-dimensional in twistor space.  The assumptions underlying the bound then amount to the statement
that you can only place a certain amount of `entropy' on a given twistor or null ray.  It would
be interesting to have some fundamental explanation for this statement.

While this construction may seem academic, it may also have some useful applications.  In particular,
one may imagine scenarios where the Bousso lightsheet may be forced to fluctuate \cite{lee}.
Such a fluctuation will generically produce some timesheet.
As long as the fluctuation is sufficiently small, so that (3) is still satisfied, then 
the covariant entropy bound will still hold.  In this sense, we feel that allowing for timesheets
in the formulation of the bound only makes the bound more robust.

Finally, we would like to mention an exciting possibility that may allow for the
construction of an entropy bound that only uses {\it topological} properties of twistor
space.  It has been known, through the work of Low and others (\cite{rob}, \cite{tim}),
that there is a relationship between {\it linking} in twistor space and causal
structure in spacetime.  To be precise, recall that a point in spacetime corresponds
to an $S^2$ in twistor space.  Then it turns out that, at least for suitably `simple'
spacetimes, two points in spacetime are causally related if and only if the corresponding
two-spheres are `linked'.  Now assume that we focus on a small causal diamond in 
the spacetime, which is globally hyerbolic with some Cauchy surface $S$, and let
$A$ be some spacelike region with boundary $B$.  
Low \cite{low} has further argued that the Bousso choice for the lightsheet of $B$ 
corresponds precisely to the future horismos of $A$ \footnote{The reader will recall
that the future horismos of $A$ is just the boundary of the Cauchy development.}. 
Put another way, in twistor space the lightsheet is just a boundary between a region
of linked spheres and a region of unlinked spheres.  Thus, the lightsheet can
be specified in terms of purely {\it topological} data in twistor space.
We would emphasize that this
can only really be true for very simple spacetimes, but it would be interesting
to see how robustly these speculations can be implemented.

\acknowledgments

AC was supported by a Director's Funded Fellowship at Los Alamos National Lab.
I would like to thank the Kavli Institute for Theoretical
Physics for hospitality while this work was supported in part
by the National Science Foundation under Grant No. PHY99-07949.
I also thank Raphael Bousso, Rob Low and Don Marolf for correspondence.

\end{multicols}

\end{document}